\documentclass[aps,prb,twocolumn,amsmath,amssymb,nofootinbib,superscriptaddress,floatfix,eqsecnum]{revtex4-1}

\usepackage{amsmath}
\usepackage{amssymb}
\usepackage{amsthm}
\usepackage[pdftex]{color} 
\usepackage{graphicx}
\usepackage{dcolumn} 
\usepackage{bm} 
\usepackage{longtable}
\usepackage{ulem}   
\newcommand{\bs}[1]{{\boldsymbol{#1}}}

\begin{document}

\title{Topological Hubbard model and its high-temperature
  quantum Hall effect}
\author{Titus Neupert} 
\affiliation{
Condensed Matter Theory Group, 
Paul Scherrer Institute, CH-5232 Villigen PSI,
Switzerland
            } 

\author{Luiz Santos} 
\affiliation{
Department of Physics, 
Harvard University, 
17 Oxford Street, 
Cambridge, Massachusetts 02138,
USA
            } 

\author{Shinsei Ryu} 
\affiliation{
Department of Physics, University of Illinois at Urbana-Champaign, 
1110 W. Green Street, 
Urbana, Illinois 61801-3080, USA}

\author{Claudio Chamon} 
\affiliation{
Physics Department, 
Boston University, 
Boston, Massachusetts 02215, USA
            } 
            
\author{Christopher Mudry} 
\affiliation{
Condensed Matter Theory Group, 
Paul Scherrer Institute, CH-5232 Villigen PSI,
Switzerland
            } 

\date{\today}

\begin{abstract}
The quintessential two-dimensional lattice model
that describes the competition between the kinetic energy
of electrons and their short-range repulsive interactions is 
the repulsive Hubbard model. We study a time-reversal symmetric variant 
of the repulsive Hubbard model defined on a planar lattice:
Whereas the interaction is unchanged,
any fully occupied band supports a quantized spin Hall effect.
We show that at 1/2 filling of this band, the ground state develops
spontaneously and simultaneously Ising ferromagnetic long-range order 
\textit{and} 
a quantized charge Hall effect when the interaction is sufficiently strong.
We ponder on the possible practical applications, beyond metrology,
that the quantized charge Hall effect might have if it could be realized at
high temperatures and without external magnetic fields in strongly
correlated materials.
\end{abstract}

\maketitle

\section{Introduction}

High-temperature superconductivity~\cite{Lee06}
and the quantum Hall effect (QHE)~\cite{Prange87} 
have been two of the central problems in condensed matter
physics of the past three decades. The former is related 
to electrons hopping on a two-dimensional (2D) 
lattice close to (but not at) half filling, while the
latter focuses on fermions in doped semiconductor heterostructures or
graphene in a high magnetic field. High-temperature superconductors
are strongly interacting systems, with the potential energy 
about an order of magnitude larger than the kinetic energy. 
In the QHE, the kinetic energy is quenched by the
external magnetic field. Moreover,
interactions are important only in
understanding the fractional QHE 
but not in understanding the integer QHE (IQHE).

The possibility that the IQHE could arise in a lattice Hamiltonian
without the Landau levels induced by a uniform magnetic field was
suggested by Haldane in 1988~\cite{Haldane88}. The essence is that,
despite the absence of an uniform magnetic field, the system still lacks
time-reversal symmetry. More recently, it was shown that the fractional QHE
could also emerge in flat topological bands when they are
partially filled%
~\cite{Neupert11a,Sheng11,Wang11a,Regnault11,Xiao11,Wang11}
(see also~\onlinecite{Tang11}). 
These recent developments point to a natural
marriage between the QHE and strongly correlated lattice systems at
a high filling fraction.

In this Letter, we study a quintessential strongly correlated lattice 2D
system but with a twist. We consider a time-reversal symmetric
fermionic Hubbard model in the limit of large on site repulsion $U$ 
compared to the bandwidth $W$ of the hopping dispersion, 
but with hopping terms yielding topologically nontrivial Bloch bands 
in that they each support a quantized spin Hall conductivity 
when fully occupied~\cite{Bernevig06a}. 
The time-reversal symmetric Hubbard model 
with a single half filled nested Bloch band has 
a charge insulating ground state that
supports antiferromagnetic long-range order~\cite{Lee06}.
In contrast, the ground state of 
our time-reversal symmetric Hubbard model 
with topologically nontrivial Bloch bands \textit{simultaneously} 
displays Ising ferromagnetic long-range order \textit{and} the IQHE 
at some commensurate filling fraction. 
The energy scales that can be
attained in lattice models are typically rather high, of the order of
atomic magnitudes, i.e., electronvolt. 
If an interacting system with topological
bands can be found so as to display the IQHE at high temperatures, it
could be of practical use, as we shall explain
after we substantiate our claims.

\section{Study of the topological Hubbard Model}

We consider spinful electrons hopping on a bipartite square lattice 
$\Lambda=A\cup B$ 
with sublattices $A$ and $B$, 
where each sublattice has 
$N:=L^{\ }_{x}\times L^{\ }_{y}$ sites.
The Hubbard Hamiltonian with repulsive interactions ($U>0$) can be written
\begin{subequations}
\label{eq: def Hamiltonian}
\begin{equation}
H:=
\sum_{\bs{k}\in\mathrm{BZ}}
c^{\dagger}_{\bs{k}}\,
\mathcal{H}^{\ }_{\bs{k}}\,
c^{\ }_{\bs{k}}
+
U
\sum_{\bs{r}}
\sum_{\alpha=A,B}
n^{\ }_{\bs{r},\uparrow,\alpha}
n^{\ }_{\bs{r},\downarrow,\alpha}.
\label{eq: def Hamiltonian a}
\end{equation}
The component $c^{\dag}_{\bs{k},\sigma,\alpha}$
of the operator-valued spinor $c^{\dag}_{\bs{k}}$ 
creates an electron with momentum 
$\bs{k}$ from the Brillouin zone (BZ) of sublattice $A$
and with spin $\sigma=\uparrow,\downarrow$,
whose Fourier transform
$c^{\dag}_{\bs{r},\sigma,\alpha}=
N^{-1/2}
\sum_{\bs{k}\in\mathrm{BZ}}
e^{-i\bs{\bs{k}}\cdot\bs{r}}\,
c^{\dag}_{\bs{k},\sigma,\alpha}$
is exclusively supported on sublattice $\alpha=A,B$. 
The $4\times4$ Hermitian matrix
$\mathcal{H}^{\ }_{\bs{k}}$
obeys the time-reversal symmetry (TRS)
\begin{equation}
\mathcal{H}^{\ }_{+\bs{k}}=
\sigma^{\ }_{2}\,
\mathcal{H}^{* }_{-\bs{k}}\,
\sigma^{\ }_{2},
\label{eq: def Hamiltonian b}
\end{equation}
and, owing to a strong intrinsic spin-orbit coupling, 
the residual spin-rotation symmetry (RSRS)
\begin{equation}
\mathcal{H}^{\ }_{+\bs{k}}=
\sigma^{\ }_{3}\,
\mathcal{H}^{\ }_{+\bs{k}}\,
\sigma^{\ }_{3}\equiv
\begin{pmatrix}
h^{(\uparrow)}_{\bs{k}}
&
0
\\
0
&
h^{(\downarrow)}_{\bs{k}}
\end{pmatrix},
\label{eq: def Hamiltonian c}
\end{equation}
where the Pauli matrices 
$\sigma^{\ }_{1}$,
$\sigma^{\ }_{2}$,
and
$\sigma^{\ }_{3}$
act on the electronic spin-1/2 degrees of freedom.
Hence, the two $2\times2$ Hermitian matrices 
$h^{(\sigma)}_{\bs{k}}$
with $\sigma=\uparrow,\downarrow$ obey
\begin{equation}
h^{(\uparrow)  }_{+\bs{k},\alpha\beta}=
h^{(\downarrow)}_{-\bs{k},\beta\alpha},
\qquad 
\forall 
\bs{k}\in\mathrm{BZ},
\qquad 
\alpha,\beta=A,B,
\label{eq: def Hamiltonian d}
\end{equation}
\end{subequations}
because of the condition of TRS%
~(\ref{eq: def Hamiltonian b}).
Finally, the operator
$
n^{\   }_{\bs{r},\sigma,\alpha}=
c^{\dag}_{\bs{r},\sigma,\alpha}
c^{\   }_{\bs{r},\sigma,\alpha}
$ 
measures the electron density on site $\bs{r}$ in sublattice $\alpha$ 
and with spin~$\sigma$.

The Hubbard Hamiltonian defined by Eq.~(\ref{eq: def Hamiltonian a})
thus has a global $\mathbb{Z}^{\ }_{2}\times U(1)$ symmetry that
arises because of the TRS~(\ref{eq: def Hamiltonian b}) and the 
RSRS~(\ref{eq: def Hamiltonian c}). 
We are going to show that TRS is spontaneously broken
while the continuous RSRS is shared by the ground state,
when this Hubbard Hamiltonian acquires
suitable topological properties.

It is the choice for the matrix elements
$h^{(\sigma)}_{\bs{k},\alpha\beta}$ 
entering the kinetic energy~(\ref{eq: def Hamiltonian c})
that endows the Hubbard Hamiltonian~(\ref{eq: def Hamiltonian a})
with topological attributes.
We choose
\begin{subequations}
\label{eq: def kinetic energy}
\begin{equation}
\begin{split}
h^{(\uparrow) }_{\bs{k},AB}=
h^{(\uparrow)*}_{\bs{k},BA}&:=
w^{\ }_{\bs{k}}
\left[
e^{-i\pi/4}
\left(
1+e^{+i(k^{\ }_{y}-k^{\ }_{x})}
\right)
\right.
\\
&\hphantom{w^{\ }_{\bs{k}}t^{\ }_1}
\left.
+
e^{+i\pi/4}
\left(
e^{-ik^{\ }_{x}}
+
e^{+ik^{\ }_{y}}
\right)
\right],\\
h^{(\uparrow)}_{\bs{k},AA}=
-
h^{(\uparrow)}_{\bs{k},BB}&:=
w^{\ }_{\bs{k}}
\left[
2t^{\ }_2(\cos\,k^{\ }_{x}-\cos\,k^{\ }_{y})
+
4\mu^{\ }_{\mathrm{s}}
\right],
\end{split}
\label{eq: def kinetic energy a}
\end{equation}
where 
\begin{equation}
w^{-1}_{\bs{k}}:=
\kappa\,
\varepsilon^{\ }_{\bs{k}}
+
(1-\kappa),\qquad 
\kappa\in[0,1],
\label{eq: def kinetic energy b}
\end{equation}
and
\begin{equation}
\varepsilon^{\ }_{\bs{k}}\!:=\!
\sqrt{
1
+
\cos
k^{\ }_{x}
\cos
k^{\ }_{y}
+
\left[
2t^{\ }_{2}
\left(
\cos
k^{\ }_{x}
-
\cos
k^{\ }_{x}
\right)
+
4\mu^{\ }_{\mathrm{s}}
\right]^2
     }.
\label{eq: def kinetic energy c}
\end{equation}
In the noninteracting limit ($U=0$), 
this model features four bands 
with two distinct two-fold degenerate dispersions 
$\pm w^{\ }_{\bs{k}}\varepsilon^{\ }_{\bs{k}}$%
~\cite{Neupert11a}.
This two-fold degeneracy is a consequence of the Kramers degeneracy
implied by the TRS~(\ref{eq: def Hamiltonian b}).
If we denote the corresponding eigenspinors  
$\chi^{\ }_{\bs{k},\sigma,\lambda}=
(\chi^{\ }_{\bs{k},\sigma,\lambda,\alpha})$,
where $\lambda=\pm$,
and choose the normalization 
$\chi^{\dag}_{\bs{k},\sigma,\lambda}\chi^{\ }_{\bs{k},\sigma,\lambda'}=
\delta^{\ }_{\lambda,\lambda'},\ 
\forall \bs{k}$,
then the kinetic energy is diagonalized by using 
the fermionic creation operators 
\begin{equation}
d^{\dag}_{\bs{k},\sigma,\lambda}:=
\sum_{\alpha=A,B}\chi^{*}_{\bs{k},\sigma,\lambda,\alpha}
c^{\dag}_{\bs{k},\sigma,\alpha},
\label{defd}
\end{equation}
as
\begin{equation}
H^{\ }_{0}:=
\sum_{\bs{k}\in\mathrm{BZ}}
\sum_{\sigma=\uparrow,\downarrow}
\sum_{\lambda=\pm}
\lambda\,
d^{\dagger}_{\bs{k},\sigma,\lambda}\,
w^{\ }_{\bs{k}}\varepsilon^{\ }_{\bs{k}}\,
d^{\ }_{\bs{k},\sigma,\lambda}.
\label{eq: def kinetic energy d}
\end{equation}
\end{subequations}
Hence, the Bloch states created by 
$d^{\dag}_{\bs{k},\sigma,\lambda}$
are generically spread on both sublattices $A$ and $B$.
We shall consider only the case in which these bands are separated 
by an energy gap, i.e., 
$|t^{\ }_2|\neq |\mu^{\ }_{\mathrm{s}}|$.
The parameter $\kappa$ controls the bandwidth of these bands. 
For $\kappa=1$, the bands are exactly flat with eigenvalues $\pm1$. 
The case $\kappa=0$ corresponds to a tight-binding model on 
the square lattice that involves only nearest-neighbor 
($|t^{\ }_{1}|= 1$) and next-nearest-neighbor hopping 
($t^{\ }_{2}$)
together with a staggered chemical potential
$\mu^{\ }_{\mathrm{s}}$ that breaks the symmetry between sublattices 
$A$ and $B$~\cite{Neupert11a}. 
For $\kappa\in(0,1]$, longer-range hopping is introduced. 
However, we stress that the Hamiltonian remains local for all 
$\kappa\in[0,1]$ since all correlation functions decay exponentially 
due to the presence of the band gap~\cite{Neupert11a}.

The topological properties of the lower pair of bands are characterized 
by their spin Chern number 
\begin{subequations}
\begin{equation}
C^{\ }_\mathrm{s}:=
\left(
C^{\ }_{\uparrow}
-
C^{\ }_{\downarrow}
\right)/2,
\end{equation}
where $C^{\ }_{\sigma}$ is 
to be computed from the orbitals of spin-$\sigma$ 
electrons according to
\begin{equation}
C^{\ }_\sigma:=
\int_{\bs{k}\in\mathrm{BZ}}
\frac{d^2\bs{k}}{2\pi i}
\bs{\nabla}^{\ }_{\bs{k}}
\wedge
\left(
\chi^{\dag}_{\bs{k},\sigma,-}
\bs{\nabla}^{\ }_{\bs{k}} 
\chi^{\ }_{\bs{k},\sigma,-}
\right).
\label{spin-resolvedChernNumber}
\end{equation}
Time-reversal symmetry implies 
$C^{\ }_\uparrow=-C^{\ }_\downarrow$ 
and therefore entails a vanishing of the total (charge) Chern number 
$C^{\ }_{\mathrm{c}}:=(C^{\ }_\uparrow+C^{\ }_\downarrow)/2$ 
of the lower bands.
The spin Chern number of the lower pair of bands is given by
\begin{equation}
C^{\ }_\mathrm{s}=\frac{1}{2}\left(
\mathrm{sgn}\,h^{(\uparrow)}_{(0,\pi),AA}
-
\mathrm{sgn}\,h^{(\uparrow)}_{(\pi,0),AA}\right).
\end{equation}
\end{subequations}
Hence, the Bloch bands are topologically trivial whenever 
$|t^{\ }_2/\mu^{\ }_{\mathrm{s}}|<1$, 
while the model at half filling exhibits the physics of a 
quantum spin Hall insulator whenever 
$|t^{\ }_2/\mu^{\ }_{\mathrm{s}}|>1$. 
In an open geometry, the spin Hall conductivity 
is quantized to the value 
$\sigma^{\mathrm{sH}}_{xy}=eC^{\ }_\mathrm{s}/(2\pi)$
where $e$ denotes the electric charge of the electron.

We now consider the system with a repulsive
Hubbard interaction $U>0$ at 
1/2 filling of the lower band (1/4 filling of the lower and upper bands), 
i.e., with 
\begin{equation}
N^{\ }_{\mathrm{e}}=L^{\ }_{x}\times L^{\ }_{y}=N
\label{eq: def commensurate filling}
\end{equation}
electrons.
In all that follows, we assume that $U$ is \textit{much smaller} 
than the gap $\Delta^{\ }_{0}$ 
induced by a strong intrinsic spin-orbit coupling
between the two pairs of bands.
If so, we can restrict the 
$N^{\ }_{\mathrm{e}}$-body Hilbert space to the Fock space
arising from the single-particle Hilbert spaces of 
the lower pair of bands. 

In the limit of flat bands $\kappa=1$ and at 
the commensurate filling fraction~(\ref{eq: def commensurate filling}),
the kinetic energy~(\ref{eq: def kinetic energy d}) 
at fixed spin polarization 
$S:=|\langle\sigma^{\ }_3\rangle|=0$, 2, ..., $N$ in units of $\hbar/2$
has a  ground state degeneracy 
\begin{equation}
\mathcal{N}_{\mathrm{gs}}=
\begin{pmatrix}
N
\\
\frac{N-|S|}{2}
\end{pmatrix}^2.
\end{equation}
The repulsive Hubbard interaction lifts this degeneracy whenever
any one of these states allows a site of $\Lambda$ to be doubly occupied
with a finite probability. The only two states with full spin polarization 
$S=N$,
\begin{subequations}
\begin{equation}
|\Psi^{\ }_{\sigma}\rangle=
\prod_{\bs{k}\in\mathrm{BZ}}
d^{\dag}_{\bs{k},\sigma,-}|0\rangle,
\qquad 
\sigma=\uparrow,\downarrow,
\label{symmbrokenstates}
\end{equation}
are immune 
to the presence of the Hubbard repulsion. 
More formally, observe that Hamiltonian 
$H-\mu N^{\ }_{\mathrm{e}}$
is a positive semidefinite operator for 
$\kappa=1$, $U>0$, and the chemical potential $\mu=-1$. 
Since
\begin{equation}
\langle\Psi^{\ }_{\sigma}|
\left(
H
+
N^{\ }_{\mathrm{e}}
\right)
|\Psi^{\ }_{\sigma}\rangle=0, 
\qquad 
\sigma=\uparrow,\downarrow,
\end{equation}
\end{subequations}
the two states \eqref{symmbrokenstates}
belong to the ground state manifold of 
$H+N^{\ }_{\mathrm{e}}$
for any $U>0$, $t^{\ }_{2}$, and $\mu^{\ }_{\mathrm{s}}$.

We are going to argue that this pair
of degenerate Ising ferromagnets spans the ground state manifold
for any $U>0$ and $|t^{\ }_{2}/\mu^{\ }_{\mathrm{s}}|\neq1$.
This is achieved by arguing that they are separated
from excited states by a many-body gap, a departure
from the usual ferromagnetism in flat bands 
when full spin-1/2 $SU(2)$ symmetry is not explicitly broken%
~\cite{Kastura10}.
First, particle-hole excitations of 
$|\Psi^{\ }_{\sigma}\rangle$ 
that keep $S=N$ fixed,
cost an energy $\Delta^{\ }_{0}>0$
and are thus gapped.
Second, we ask whether 
excitations of $|\Psi^{\ }_{\sigma}\rangle$ 
that flip one spin ($S=N-2$) are gapped as well.
Any such state can be written as
\begin{subequations}
\begin{equation}
|\Phi^{\ }_{\sigma,\bs{Q}}\rangle=
\sum_{\bs{k}\in\mathrm{BZ}}A^{(\bs{Q})}_{\bs{k}}
d^{\dag}_{\bs{k}+\bs{Q},\bar{\sigma},-}
d^{\   }_{\bs{k},\sigma,-}
|\Psi^{\ }_{\sigma}\rangle,
\qquad \sigma=\uparrow,\downarrow,
\end{equation}
where the center of mass momentum $\bs{Q}$ is a good quantum number and thus 
$\langle\Phi^{\ }_{\sigma,\bs{Q}}|\Phi_{\sigma,\bs{Q}'}\rangle=
 \delta^{\ }_{\bs{Q},\bs{Q}'}$
if the normalization 
$\sum^{\ }_{\bs{k}\in\mathrm{BZ}}
A^{(\bs{Q})*}_{\bs{k}}A^{(\bs{Q})}_{\bs{k}}=1$ 
is imposed. One verifies that (see Appendix)
\begin{equation}
\begin{split}
&
\langle\Phi^{\ }_{\sigma,\bs{Q}}|
\left(
H
+
N^{\ }_{\mathrm{e}}
\right)
|\Phi^{\ }_{\sigma,\bs{Q}}\rangle
\\
&\quad=
U-\frac{U}{N}\sum_{\alpha}
\left| 
\sum^{\ }_{\bs{k}\in\mathrm{BZ}}
A^{(\bs{Q})}_{\bs{k}}
\chi^{\ }_{-\bs{k}-\bs{Q},\sigma,-,\alpha}
\chi^{\ }_{\bs{k},\sigma,-,\alpha}
\right|^2,
\label{matrixElement}
\end{split}
\end{equation}
\end{subequations}
where the lowest energy state with one spin flipped is characterized by the 
$A^{(\bs{Q})}_{\bs{k}}$ that minimizes Eq.~\eqref{matrixElement} 
while satisfying the normalization condition. 
For example, 
if the single-particle orbitals are fully sublattice polarized, 
e.g., $\chi^{\dag}_{\bs{k},\sigma,-}\propto(1,0)$ 
(topologically trivial), the choice $A^{(\bs{Q})}_{\bs{k}}=N^{-1/2}$ 
minimizes Eq.~\eqref{matrixElement} with the right-hand side 
equal to zero. Hence, the fully spin-polarized state 
$|\Psi^{\ }_{\sigma}\rangle$ 
is a \emph{gapless} ground state in this case. 
On the other hand, let us assume that
\begin{equation}
\chi^{\dagger}_{\bs{k},\sigma,-}\not\propto(1,0)
\quad
\text{and}
\quad
\chi^{\dagger}_{\bs{k},\sigma,-}\not\propto(0,1)
\label{nonpolarizedcondition}
\end{equation}
holds almost everywhere in the BZ, i.e., up to a set of measure zero.
In the thermodynamic limit, where the sum over $\bs{k}$ becomes an integral,
this delivers from Eq.~\eqref{matrixElement} the \emph{strict} inequality (see Appendix)
\begin{equation}
\left\langle\Phi^{\ }_{\sigma,\bs{Q}}\left|
\left(
H
+
N^{\ }_{\mathrm{e}}
\right)
\right|\Phi^{\ }_{\sigma,\bs{Q}}\right\rangle
>
0.
\label{gapproof}
\end{equation}
Hence, assumption~\eqref{nonpolarizedcondition} 
is sufficient to show that the spin-polarized state 
$|\Psi^{\ }_{\sigma}\rangle$ 
is a \emph{gapped} ground state of the Hamiltonian 
with flat bands in the thermodynamic limit, 
provided that one also assumes that the lowest energy states 
with more than one spin flipped
are higher in energy than those with one spin flipped.

Ruling out the possibility
that states with many spin flips have lower energies than
states with few spin flips relative to the Ising ferromagnetic
ground state is plausible in the regime when
the intrinsic spin-orbit coupling generates 
the largest energy scale
($\Delta^{\ }_{0}\gg U$).

Equation~\eqref{nonpolarizedcondition} 
is a reasonable assumption when the Bloch states 
stem from a band with nonzero (spin) Chern number, 
since the spinor 
$\chi^{\ }_{\bs{k},\sigma}$ 
maps out the entire surface of the unit sphere as $\bs{k}$ 
takes values in the BZ. 
The assumption underlying Eq.~\eqref{nonpolarizedcondition}
can also be understood by constructing the
Wannier wavefunctions,  centered at the lattice point $\bs{z}$,
of the lowest energy band with spin $\sigma$ and Chern number:
$C^{\ }_{\sigma}$
\begin{subequations}
\begin{equation}
\psi^{\ }_{\bs{z},\bs{r},\sigma,-,\alpha}:=
\frac{1}{N}\,\sum_{\bs{k}\in\rm{BZ}}\,
e^{i\bs{k}\cdot(\bs{r}-\bs{z})}\,
\chi^{\ }_{\bs{k},\sigma,-,\alpha}.
\label{eq: def Wannier wave fct}
\end{equation}
The gauge-invariant part of their spread functional~\cite{Thonhauser06}
satisfies
\begin{equation}
\label{eq: lower bound spread}
\begin{split}
&
\langle\,\psi^{\ }_{\bs{0},\sigma,-}\,|\,\bs{r}^{2}|\psi^{\ }_{\bs{0},\sigma,-}\,\rangle
-
\sum_{\bs{z}}\,
|\,\langle\,\psi^{\ }_{\bs{0},\sigma,-}\,|\,
\bs{r}
|\psi^{\ }_{\bs{z},\sigma,-}\,\rangle\,|^{2}
\\
&\qquad\geq
|C^{\ }_{\sigma}|\,
\mathcal{A}^{\ }_{c}/(2\pi),
\end{split}
\end{equation}
\end{subequations}
where   
$\mathcal{A}^{\ }_{c}$ denotes the
area of the unit cell. This inequality relates the Chern number of the
band and the ``minimum width'' of the Wannier states (see Appendix).
In particular, in the nontopological phase, 
one can imagine a limit
in which the wavefunction is entirely localized on a given sublattice, while 
the nonzero Chern number in the topological phase implies that the
Wannier wavefunction has amplitudes on both sublattices. 

\begin{figure}
\includegraphics[width=0.48\textwidth]{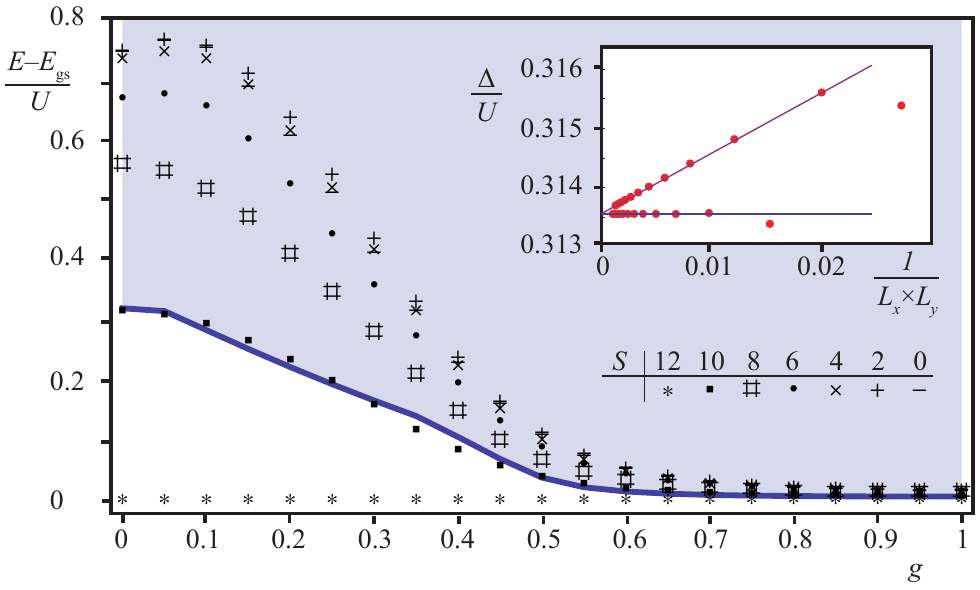}
\caption{
Numerical exact diagonalization results for flat bands $\kappa=1$ 
at the commensurate filling fraction%
~(\ref{eq: def commensurate filling}).
Markers show the energy of the lowest state in different sectors of 
total spin $S$ (in units of $\hbar/2$) 
measured with respect to the ground state energy for 
$L^{\ }_{x}=3,\ L^{\ }_{y}=4$. 
Here, 
$g:=(2/\pi)\mathrm{arctan}|\mu^{\ }_{\mathrm{s}}/t^{\ }_{2}|$ 
so that $g>0.5$ and $g<0.5$ 
correspond to the trivial and topological single-particle bands, 
respectively. Since there is only one state in the fully polarized sector 
$|S|=12$, the difference between the asterisks and the squares is 
the many-body excitation gap $\Delta(g)$.
The thick blue line shows the extrapolation of $\Delta(g)$ 
to the thermodynamic limit. In the inset, exact diagonalization in
the sector with one spin flipped away from the fully polarized sector
is presented for $\mu^{\ }_{\mathrm{s}}=0$, $t^{\ }_{2}=1/\sqrt{2}$
and $L^{\ }_{x}=L^{\ }_{y}$ ranging from $6$ to $30$. 
The straight lines are a guide to the eye and make evident an even-odd 
effect in $L^{\ }_{x}=L^{\ }_{y}$. 
Deep in the topologically nontrivial regime
$g\ll0.5$, we observe a sizable $\Delta(g\ll0.5)$.
The topologically trivial regime $g>0.5$ is also characterized by a gap
$\Delta(g>0.5)$
in the sector with one spin flipped away from the fully polarized sector,
however this gap is much smaller than $\Delta(g\ll0.5)$.
We refer the reader to the Appendix for a discussion of
the regime $\Delta(g>0.5)$.
\label{fig:1} 
        }
\end{figure}

While Eq.~\eqref{gapproof} is strongly suggestive of
the existence of a many-body gap $\Delta$, 
it does not provide information
about its size.
To quantify $\Delta$, we diagonalized the model%
~\eqref{eq: def Hamiltonian} 
exactly numerically in the 
limit of flat bands $\kappa=1$ at 
the commensurate filling fraction~(\ref{eq: def commensurate filling}).
We varied the ratio $|t^{\ }_2/\mu^{\ }_{\mathrm{s}}|$, 
keeping $t^{2}_2+\mu^{2}_{\mathrm{s}}=1/2$ 
constant to drive the system from the topological to the trivial phase.
The results are shown in Fig.~\ref{fig:1}.
First, they support the assumption that all states with more than one spin 
flipped are higher in energy than the many-body one-spin-flipped gap
provided $|t^{\ }_{2}/\mu^{\ }_{\mathrm{s}}|>1$.
Second, we find $\Delta\approx 0.3 U$ as an extrapolation to 
the thermodynamic limit 
for $\mu^{\ }_{\mathrm{s}}=0$ deep in the topological phase, 
while $\Delta$ monotonically decreases toward 
a much smaller nonvanishing value
for $t^{\ }_{2}=0$ in the topologically trivial phase 
set by the unit of energy $|t^{\ }_{1}|=1$.
Finally, it should be noted that neglecting the states from the upper band
of the noninteracting Hamiltonian
delivers the correct excitation many-body gap $\Delta$ 
not only in the aforementioned 
limit $U\ll \Delta^{\ }_0$,
but also under the weaker condition $\Delta<\Delta^{\ }_0$,
if the limit of flat bands is taken.

Deep in the topologically nontrivial regime 
$|t^{\ }_{2}/\mu^{\ }_{\mathrm{s}}|\gg1$,
the states 
$|\Psi^{\ }_{\sigma}\rangle$ 
and 
$|\Psi^{\ }_{\bar{\sigma}}\rangle$ 
are degenerate ground states related by TRS
for any finite
$N$. They are separated from their excitations by a gap that survives the
thermodynamic limit $N\to\infty$. Spontaneous breaking of TRS
takes place in the thermodynamic limit $N\to\infty$ by
selecting the ground state to be $|\Psi^{\ }_{\uparrow}\rangle$, say.
It is then meaningful to discuss the quantized electromagnetic response of 
$|\Psi^{\ }_{\uparrow}\rangle$,
since TRS is spontaneously broken.
The transverse charge response $\sigma^{\mathrm{H}}_{xy}$
of $|\Psi^{\ }_{\uparrow}\rangle$
is proportional to the many-body Chern number 
$C^{\ }_{|\Psi^{\ }_{\uparrow}\rangle}$. 
The latter takes into account the occupation of the 
Bloch states~\cite{Neupert11a}. 
Since all Bloch states
of the lower band with spin $\sigma$ are occupied in 
$|\Psi^{\ }_{\uparrow}\rangle$, 
while all Bloch states with spin $\downarrow$ are empty, 
$C^{\ }_{|\Psi^{\ }_{\uparrow}\rangle}\equiv C^{\ }_{\uparrow}$. 
Hence, the ground state has the quantized Hall response 
\begin{equation}
|\sigma^{\mathrm{H}}_{xy}|=|C^{\ }_{\uparrow}|\times e^{2}/h=e^{2}/h.
\end{equation}

\begin{figure}
\includegraphics[width=0.44\textwidth]{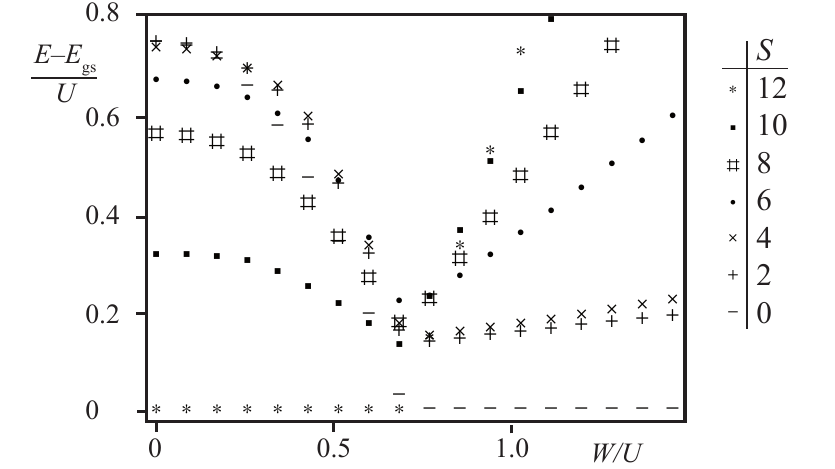}
\caption{
(Color online) 
Numerical exact diagonalization results at 
the commensurate filling fraction~(\ref{eq: def commensurate filling})
as a function of the bandwidth $W$
for $L^{\ }_{x}=3,\ L^{\ }_{y}=4$. Plotted is the energy of the lowest state 
in different sectors of total spin $S$ (in units of $\hbar/2$) 
measured with respect to the ground state energy in the topological phase with 
$\mu^{\ }_{\mathrm{s}}=0$, $t^{\ }_{2}=1/\sqrt{2}$. 
The ground state is gapped and fully spin-polarized for $W/U<0.7$, 
while it is unpolarized for $W/U>0.7$ (see Appendix).
\label{fig:2} 
        }
\end{figure}

Remarkably, the selection by the repulsive Hubbard interaction
of a ground state supporting \textit{simultaneously} 
Ising ferromagnetism \textit{and} the IQHE is 
robust to a sizable bandwidth as is suggested by
numerical exact diagonalization.
As shown in Fig.~\eqref{fig:2}, the fully spin-polarized state 
$|\Psi^{\ }_{\sigma}\rangle$ remains the gapped ground state 
of the system up to a bandwidth $W/U\approx 0.7$.

\section{Practical applications}

So what is it good for, a material with a QHE at room temperature 
without applied external magnetic fields besides metrology%
~\cite{Klitzing05,Matthews05}?
First, we recall that the quantization of the Hall resistance
and the accompanying vanishing of the longitudinal resistance  
is exact only at zero temperature. The longitudinal resistance 
increases exponentially fast with increasing temperature%
~\cite{Matthews05}. 
However, if a QHE with gaps of the order of hundreds of meV or even eV 
scales could arise in a strongly correlated lattice material, 
exceptionally low resistivities could be attained. 
The resistance of a Hall bar depends
on its aspect ratio and the Hall angle
$\delta=\arctan(\rho^{\ }_{xy}/\rho^{\ }_{xx})$~\cite{Rendell81}, but for
long systems (``wires'') near the quantized regime, the longitudinal
resistance scales as $R^{\ }_{xx}=L/W\,\rho_{xx}$, and the 2D
resistivity $\rho^{\ }_{xx}\sim R^{\ }_{K} \;e^{-\Delta/T}$, 
where $\Delta$ is the excitation gap. 
For gaps of the order of $100\,$meV to $1\,$eV, one
would obtain room temperature 2D resistivities from 
$\rho^{\ }_{xx}\sim 10^3\,\Omega$ to $\rho^{\ }_{xx}\sim 10^{-13}\,\Omega$,
respectively. Obviously the exponential behavior is responsible for
this gigantic range. Small as they are, these are not perfect
conductors. For a benchmark, we consider the conductivity of copper at room
temperature per atomic layer. Using the value for
the 3D resistivity of copper at 20$^\circ$C of $\rho^{3D}_{\rm Cu}=
1.68\times 10^{-8}\, \Omega\,$m ~\cite{Matula79} and that the lattice
parameter for FCC lattice is 3.61\AA, we obtain $\rho^{2D}_{\rm Cu}=
93.3\,\Omega$. Therefore, for gaps above $\Delta \approx .2\,$eV, the
Hall system starts to be better conducting than copper at room
temperature, and for $\Delta \approx .3\,$eV it is already almost
3 orders of magnitude better conducting than copper.

The RSRS~(\ref{eq: def Hamiltonian c})
is not exact in practice. For example, a Rashba spin-orbit
coupling violates this RSRS. However, our analysis of transport
at room temperature still applies provided the characteristic
energy scale associated to the breaking of the
RSRS~(\ref{eq: def Hamiltonian c})
is much smaller than the largest energy scale $\Delta^{\ }_{0}$
induced by the intrinsic spin-orbit coupling.
Materials that realize 
a 2D $\mathbb{Z}^{\ }_{2}$ topological band insulator%
~\cite{Kane05a,Kane05b} 
with a band gap $\Delta^{\ }_{0}$
are thus candidates to realize a QHE at room temperature
if (i) the band gap is larger than
the correlation energy  and (ii) the chemical potential
can be tuned to half-filling of (iii) a reasonably flat valence band.
HgTe quantum wells with an inverted band structure 
realize a 2D $\mathbb{Z}^{\ }_{2}$ topological band insulator
with a small Rashba coupling%
~\cite{Bernevig06b,Konig07}.
The design of a material with the functionalities (i)-(iii)
has been proposed in Ref.~\onlinecite{Xiao11} .
Cold atoms trapped in an optical honeycomb lattice%
~\cite{Zhu07,Tarruell11}
might offer an alternative to realizing
the topological Hubbard model discussed in this Letter.

We would like to close by mentioning that examples such as the
topological Hubbard model discussed in this Letter, 
as well as lattice models displaying the  fractional quantum Hall effect studied in
Refs.~\onlinecite{Neupert11a,Sheng11,Wang11a,Regnault11}, could
serve as benchmarks for numerical methods of fermionic models in
2D such as dynamical mean-field theory and
methods based on tensor product states~\cite{methods}. 
In contrast to the single-band repulsive Hubbard model, 
for which little is known exactly at fractional filling, 
the topological Hubbard model~(\ref{eq: def Hamiltonian}),
because of the nonvanishing Chern numbers of its bands, leads to much
better understood (topological) ground states. It can thus serve
as a yardstick for the performance of these methods.

This work was supported in part by DOE Grant No. DEFG02-06ER46316
and by the Swiss National Science Foundation. 
\appendix

\section{
Intermediary steps
        }

\subsection{Derivation of Eq.~(\ref{matrixElement})}

For the limit of flat bands $\kappa=1$, 
we are going to compute the expectation value 
\begin{subequations}
\begin{equation}
\langle\Phi^{\ }_{\sigma,\bs{Q}}|
\left(
H
+
N^{\ }_{\mathrm{e}}
\right)
|\Phi^{\ }_{\sigma,\bs{Q}}\rangle
\end{equation}
in the $N^{\ }_{\mathrm{e}}$-many-body state 
\begin{equation}
|\Phi^{\ }_{\sigma,\bs{Q}}\rangle=
\sum_{\bs{k}\in\mathrm{BZ}}A^{(\bs{Q})}_{\bs{k}}
d^{\dag}_{\bs{k}+\bs{Q},\bar{\sigma},-}
d^{\   }_{\bs{k},\sigma,-}
|\Psi^{\ }_{\sigma}\rangle,
\qquad \sigma=\uparrow,\downarrow,
\end{equation}
which has one spin flipped as compared to 
the 
-- 
up to time-reversal symmetry and the global gauge phase factor
-- 
unique normalized $N^{\ }_{\mathrm{e}}$-many-body
state with full spin polarization
\begin{equation}
|\Psi^{\ }_{\sigma}\rangle=
\prod_{\bs{k}\in\mathrm{BZ}}
d^{\dag}_{\bs{k},\sigma,-}|0\rangle,
\qquad 
\sigma=\uparrow,\downarrow.
\end{equation}
Here, $\bs{Q}$ denotes the center of mass momentum 
and the orthonormalization condition 
\begin{equation}
\langle\Phi^{\ }_{\sigma,\bs{Q}}|\Phi_{\sigma,\bs{Q}'}\rangle=
 \delta^{\ }_{\bs{Q},\bs{Q}'}
\end{equation}
enforces the normalization 
\begin{equation}
\sum^{\ }_{\bs{k}\in\mathrm{BZ}}
A^{(\bs{Q})*}_{\bs{k}}A^{(\bs{Q})}_{\bs{k}}=1.
\label{NormA}
\end{equation}
\end{subequations}

\medskip
\begin{widetext}
We rewrite $|\Phi_{\sigma,\bs{Q}}\rangle$ as
\begin{subequations}
\begin{equation}
\begin{split}
|\Phi^{\ }_{\sigma,\bs{Q}}\rangle
=&\,
\sum_{\bs{k}\in\mathrm{BZ}}
A^{(\bs{Q})}_{\bs{k}}
\left(
\sum_{\alpha=A,B}
\chi^{*}_{\bs{k}+\bs{Q},\bar{\sigma},-,\alpha}\,
c^{\dag}_{\bs{k}+\bs{Q},\bar{\sigma},\alpha}
\right)
\left(
\sum_{\beta=A,B}\chi^{\ }_{\bs{k},\sigma,-,\beta}\,
c^{\ }_{\bs{k},\sigma,\beta}
\right)
|\Psi^{\ }_{\sigma}\rangle
\\
=&\,
\sum_{\alpha,\beta=A,B}
\sum_{\bs{r},\bs{r}'\in A}
\left[
\sum_{\bs{k}\in\mathrm{BZ}}
\frac{A^{(\bs{Q})}_{\bs{k}}}{N}
\chi^{*}_{\bs{k}+\bs{Q},\bar{\sigma},-,\alpha}
\chi^{\ }_{\bs{k},\sigma,-,\alpha}
e^{-i(\bs{k}+\bs{Q})\cdot\bs{r}}
e^{+i\bs{k}\cdot\bs{r}'}
\right]
c^{\dag}_{\bs{r},\bar{\sigma},\alpha}
c^{\   }_{\bs{r}',\sigma,     \beta }
|\Psi^{\ }_{\sigma}\rangle
\\
\equiv&\,
\sum_{\alpha,\beta=A,B}
\sum_{\bs{r},\bs{r}'\in A}
M^{(\sigma)}_{\bs{r},\alpha,\bs{r'},\beta}\,
c^{\dag}_{\bs{r},\bar{\sigma},\alpha}
c^{\   }_{\bs{r}',\sigma,     \beta }
|\Psi^{\ }_{\sigma}\rangle,
\end{split}
\end{equation}
where we have introduced the short-hand notation
\begin{equation}
M^{(\sigma)}_{\bs{r},\alpha,\bs{r'},\beta}:=
\sum_{\bs{k}\in\mathrm{BZ}}
\frac{A^{(\bs{Q})}_{\bs{k}}}{N}
\chi^{*}_{\bs{k}+\bs{Q},\bar{\sigma},-,\alpha}
\chi^{\ }_{\bs{k},\sigma,-,\beta}
e^{-i(\bs{k}+\bs{Q})\cdot\bs{r} }
e^{+i\bs{k} \cdot\bs{r}'}.
\label{eq: def M}
\end{equation}
\end{subequations}
in terms of which
\begin{equation}
\begin{split}
\langle\Phi^{\ }_{\uparrow,\bs{Q}}|
\left(
H
+
N^{\ }_{\mathrm{e}}
\right)
|\Phi^{\ }_{\uparrow,\bs{Q}}\rangle
=&\,
U
\sum_{\tilde{\bs{r}}\in A}
\sum_{\delta=A,B}
\langle\Phi^{\ }_{\uparrow,\bs{Q}}|
n^{\ }_{\tilde{\bs{r}},\uparrow,\delta}n^{\ }_{\tilde{\bs{r}},\downarrow,\delta}
|\Phi^{\ }_{\uparrow,\bs{Q}}\rangle
\\
=&\,
U
\sum_{\tilde{\bs{r}},\bs{r},\bs{r}'\in A}
\sum_{\alpha,\beta,\delta=A,B}
\left|
M^{(\uparrow)}_{\bs{r},\alpha,\bs{r'},\beta}
\right|^2
\langle\Psi^{\ }_{\uparrow}|
c^{\dagger}_{\bs{r}',\uparrow,\beta}
c^{\ }_{\bs{r},\downarrow,\alpha}
n^{\ }_{\tilde{\bs{r}},\uparrow,\delta}
n^{\ }_{\tilde{\bs{r}},\downarrow,\delta}
c^{\dag}_{\bs{r},\downarrow,\alpha}
c^{\ }_{\bs{r}',\uparrow,\beta}
|\Psi^{\ }_{\uparrow}\rangle
\\
=&\,
U
\sum_{\tilde{\bs{r}},\bs{r},\bs{r}'\in A}
\sum_{\alpha,\beta,\delta=A,B}
\left|
M^{(\uparrow)}_{\bs{r},\alpha,\bs{r'},\beta}
\right|^2
(1-\delta^{\ }_{\bs{r},\bs{r}'}\delta^{\ }_{\alpha,\beta})
\delta^{\ }_{\tilde{\bs{r}},\bs{r}}\delta^{\ }_{\delta,\alpha}
\\
=&\,
U
\sum_{\bs{r},\bs{r}'\in A}
\sum_{\alpha,\beta=A,B}
\left|
M^{(\uparrow)}_{\bs{r},\alpha,\bs{r'},\beta}
\right|^2
-
U
\sum_{\bs{r}\in A}
\sum_{\alpha=A,B}
\left|
M^{(\uparrow)}_{\bs{r},\alpha,\bs{r},\alpha}
\right|^2.
\end{split}
\label{Sep two terms}
\end{equation}

Using the definition~(\ref{eq: def M}),
the first term becomes
\begin{equation}
\begin{split}
\sum_{\bs{r},\bs{r}'\in A}
\sum_{\alpha,\beta=A,B}
\left|
M^{(\uparrow)}_{\bs{r},\alpha,\bs{r'},\beta}
\right|^2
=&\,
\sum_{\bs{r},\bs{r}'\in A}
\sum_{\alpha,\beta=A,B}
\sum_{\bs{k},\bs{k}'\in\mathrm{BZ}}
\frac{A^{(\bs{Q})}_{\bs{k}}A^{(\bs{Q})*}_{\bs{k}'}}{N^2}
\chi^{*}_{\bs{k}+\bs{Q},\downarrow,-,\beta}
\chi^{\ }_{\bs{k},\uparrow,-,\alpha}
\chi^{\ }_{\bs{k}'+\bs{Q},\downarrow,-,\beta}
\chi^{*}_{\bs{k}',\uparrow,-,\alpha}
e^{-i(\bs{k}-\bs{k}')\cdot(\bs{r}-\bs{r}')}
\\
=&\,
\sum_{\alpha,\beta=A,B}
\sum_{\bs{k},\bs{k}'\in\mathrm{BZ}}
\frac{A^{(\bs{Q})}_{\bs{k}}A^{(\bs{Q})*}_{\bs{k}'}}{N^2}
\chi^{*}_{\bs{k}+\bs{Q},\downarrow,-,\beta}
\chi^{\ }_{\bs{k},\uparrow,-,\alpha}
\chi^{\ }_{\bs{k}'+\bs{Q},\downarrow,-,\beta}
\chi^{*}_{\bs{k}',\uparrow,-,\alpha}
N^2\delta^{\ }_{\bs{k},\bs{k}'}
\\
=&\,
\sum_{\bs{k},\in\mathrm{BZ}}
A^{(\bs{Q})}_{\bs{k}}A^{(\bs{Q})*}_{\bs{k}}
\left(
\sum_{\beta=A,B}
\chi^{*}_{\bs{k}+\bs{Q},\downarrow,-,\beta}
\chi^{\ }_{\bs{k}+\bs{Q},\downarrow,-,\beta}
\right)
\left(
\sum_{\alpha=A,B}
\chi^{\ }_{\bs{k},\uparrow,-,\alpha}
\chi^{*}_{\bs{k},\uparrow,-,\alpha}
\right)
\\
=&\,
\sum_{\bs{k},\in\mathrm{BZ}}
A^{(\bs{Q})}_{\bs{k}}A^{(\bs{Q})*}_{\bs{k}}
\\
=&\,
1,
\end{split}
\end{equation}
where both the normalization conditions on $A^{(\bs{Q})}_{\bs{k}}$ in 
Eq.~\eqref{NormA}
and on $\chi^{\ }_{\bs{k},\sigma,-,\alpha}$ 
have been used in the last two lines.
The second term in Eq.~\eqref{Sep two terms} becomes
\begin{equation}
\begin{split}
\sum_{\bs{r}\in A}
\sum_{\alpha=A,B}
\left|
M^{(\uparrow)}_{\bs{r},\alpha,\bs{r},\alpha}
\right|^2
=&\,
\sum_{\bs{r}\in A}
\sum_{\alpha=A,B}
\sum_{\bs{k},\bs{k}'\in\mathrm{BZ}}
\frac{A^{(\bs{Q})}_{\bs{k}}A^{(\bs{Q})*}_{\bs{k}'}}{N^2}
\chi^{*}_{\bs{k}+\bs{Q},\downarrow,-,\alpha}
\chi^{\ }_{\bs{k},\uparrow,-,\alpha}
\chi^{\ }_{\bs{k}'+\bs{Q},\downarrow,-,\alpha}
\chi^{*}_{\bs{k}',\uparrow,-,\alpha}
\\
=&\,
\sum_{\alpha=A,B}
\sum_{\bs{k},\bs{k}'\in\mathrm{BZ}}
\frac{A^{(\bs{Q})}_{\bs{k}}A^{(\bs{Q})*}_{\bs{k}'}}{N}
\chi^{*}_{\bs{k}+\bs{Q},\downarrow,-,\alpha}
\chi^{\ }_{\bs{k},\uparrow,-,\alpha}
\chi^{\ }_{\bs{k}'+\bs{Q},\downarrow,-,\alpha}
\chi^{*}_{\bs{k}',\uparrow,-,\alpha}
\\
=&\,
\frac{1}{N}
\sum_{\alpha=A,B}
\left|
\sum_{\bs{k}\in\mathrm{BZ}}
A^{(\bs{Q})}_{\bs{k}}
\chi^{*}_{\bs{k}+\bs{Q},\downarrow,-,\alpha}
\chi^{\ }_{\bs{k},\uparrow,-,\alpha}
\right|^2
\\
=&\,
\frac{1}{N}
\sum_{\alpha=A,B}
\left|
\sum_{\bs{k}\in\mathrm{BZ}}
A^{(\bs{Q})}_{\bs{k}}
\chi^{\ }_{-\bs{k}-\bs{Q},\uparrow,-,\alpha}
\chi^{\ }_{\bs{k},\uparrow,-,\alpha}
\right|^2,
\end{split}
\end{equation}
where we used the identity 
$\chi^{\ }_{\bs{k},\sigma,\lambda,\alpha}=
\chi^{*}_{-\bs{k},\bar{\sigma},\lambda,\alpha}$
which follows from the time-reversal symmetry of the Hamiltonian.
Putting everything together, we obtain Eq.~(\ref{matrixElement}) 
of the Letter,
\begin{equation}
\langle\Phi^{\ }_{\sigma,\bs{Q}}|
\left(
H
+
N^{\ }_{\mathrm{e}}
\right)
|\Phi^{\ }_{\sigma,\bs{Q}}\rangle
=
U-\frac{U}{N}\sum_{\alpha=A,B}
\left| 
\sum^{\ }_{\bs{k}\in\mathrm{BZ}}
A^{(\bs{Q})}_{\bs{k}}
\chi^{\ }_{-\bs{k}-\bs{Q},\sigma,-,\alpha}
\chi^{\ }_{\bs{k},\sigma,-,\alpha}
\right|^2.
\end{equation}

\end{widetext}

\subsection{Derivation of Eq.~(\ref{gapproof})}

Before starting with the derivation of the inequality Eq.~(\ref{gapproof}), 
let us establish the following inequality
\begin{equation}
1\geq a b c d +\sqrt{1-a^2}\sqrt{1-b^2}\sqrt{1-c^2}\sqrt{1-d^2}
\label{inequality}
\end{equation}
for $a,b,c,d\in [0,1]$. In particular, 
the equality only holds when either $a=b=c=d=0$ or $a=b=c=d=1$.
To show this, one can rewrite the inequality 
in terms of trigonometric functions with angles
$\alpha,\beta,\gamma,\delta\in [0,\pi/2]$
\begin{equation}
\begin{split}
1\geq &\,
\sin\, \alpha\, 
\sin\, \beta\,
\sin\, \gamma\,
\sin\, \delta\,
+
\cos\, \alpha\, 
\cos\, \beta\,
\cos\, \gamma\,
\cos\, \delta\,
\\
=&\,
\frac{1}{2}
\left[
\cos(\alpha+\beta)
\cos(\gamma+\delta)
+
\cos(\alpha-\beta)
\cos(\gamma-\delta)
\right].
\end{split}
\end{equation}
Hence the inequality holds and the equality is true only if either 
$\alpha=\beta=\gamma=\delta=0$ or
$\alpha=\beta=\gamma=\delta=\pi/2$ as announced.

We will now show that in the thermodynamic limit the \emph{strict} inequality
\begin{equation}
U-\frac{U}{N}\sum_{\alpha=A,B}
\left| 
\sum^{\ }_{\bs{k}\in\mathrm{BZ}}
A^{(\bs{Q})}_{\bs{k}}
\chi^{\ }_{-\bs{k}-\bs{Q},\sigma,-,\alpha}
\chi^{\ }_{\bs{k},\sigma,-,\alpha}
\right|^2>0
\end{equation}
holds under the assumption that 
Eq.~(\ref{nonpolarizedcondition})
is satisfied almost everywhere in the BZ, i.e., up to a set of measure zero.

\medskip
\begin{widetext}

We rewrite, for very large $N$, 
\begin{equation}
\begin{split}
&
\frac{U}{N}\sum_{\alpha=A,B}
\left| 
\sum^{\ }_{\bs{k}\in\mathrm{BZ}}
A^{(\bs{Q})}_{\bs{k}}
\chi^{\ }_{-\bs{k}-\bs{Q},\sigma,-,\alpha}
\chi^{\ }_{\bs{k},\sigma,-,\alpha}
\right|^2\\
&\qquad\rightarrow
N U
\int \frac{d^2\bs{k}}{(2\pi)^2}
\int \frac{d^2\bs{k}'}{(2\pi)^2}
A^{(\bs{Q}) }_{\bs{k} }
A^{(\bs{Q})*}_{\bs{k}'}
\sum_{\alpha=A,B}
\chi^{\ }_{-\bs{k}-\bs{Q},\sigma,-,\alpha}
\chi^{\ }_{\bs{k},\sigma,-,\alpha}
\chi^{*}_{-\bs{k}'-\bs{Q},\sigma,-,\alpha}
\chi^{*}_{\bs{k}',\sigma,-,\alpha}
\\
&\qquad
\leq
N U
\int \frac{d^2\bs{k}}{(2\pi)^2}
\int \frac{d^2\bs{k}'}{(2\pi)^2}
\left|A^{(\bs{Q}) }_{\bs{k} }\right|
\left|A^{(\bs{Q})*}_{\bs{k}'}\right|
\sum_{\alpha=A,B}
\left|\chi^{\ }_{-\bs{k}-\bs{Q},\sigma,-,\alpha}\right|
\left|\chi^{\ }_{\bs{k},\sigma,-,\alpha}\right|
\left|\chi^{*}_{-\bs{k}'-\bs{Q},\sigma,-,\alpha}\right|
\left|\chi^{*}_{\bs{k}',\sigma,-,\alpha}\right|\\
&\qquad
<
N U
\int \frac{d^2\bs{k}}{(2\pi)^2}
\int \frac{d^2\bs{k}'}{(2\pi)^2}
\left|A^{(\bs{Q}) }_{\bs{k} }\right|
\left|A^{(\bs{Q})*}_{\bs{k}'}\right|\\
&\qquad
\leq
N U
\int \frac{d^2\bs{k}}{(2\pi)^2}
\left|A^{(\bs{Q})}_{\bs{k}}\right|^2\\
&\qquad
=U.
\end{split}
\label{eq: chain inequalities}
\end{equation}
Due to the normalization of the eigenspinors,
the  assumption~\eqref{nonpolarizedcondition}, 
and the inequality~\eqref{inequality},
\begin{equation}
\sum_{\alpha=A,B}
\left|\chi^{\ }_{-\bs{k}-\bs{Q},\sigma,-,\alpha}\right|
\left|\chi^{\ }_{\bs{k},\sigma,-,\alpha}\right|
\left|\chi^{*}_{-\bs{k}'-\bs{Q},\sigma,-,\alpha}\right|
\left|\chi^{*}_{\bs{k}',\sigma,-,\alpha}\right|
<1
\end{equation}
holds for almost all $\bs{k},\ \bs{k}'$. 
This allows the use of the strict inequality in the third-last inequality
from the chain of inequalities~(\ref{eq: chain inequalities}).

\medskip
\end{widetext}

The penultimate inequality
from the chain of inequalities~(\ref{eq: chain inequalities})
is a consequence of H\"older's inequality
\begin{equation}
\int \frac{d^2\bs{k}}{(2\pi)^2}
\left|A^{(\bs{Q})}_{\bs{k}}\right|
\leq
\left(
\int \frac{d^2\bs{k}}{(2\pi)^2}
\left|A^{(\bs{Q})}_{\bs{k}}\right|^2
\right)^{1/2}.
\end{equation}
The last inequality 
in the chain of inequalities~(\ref{eq: chain inequalities})
follows from the representation 
\begin{equation}
\int \frac{d^2\bs{k}}{(2\pi)^2}
\left|A^{(\bs{Q})}_{\bs{k}}\right|^2=
\frac{1}{N}
\end{equation}
of the normalization~\eqref{NormA} in the thermodynamic limit.
In summary,
we have shown that
\begin{equation}
\left\langle\Phi^{\ }_{\sigma,\bs{Q}}
\left|
\left(
H
+
N^{\ }_{\mathrm{e}}
\right)
\right|\Phi^{\ }_{\sigma,\bs{Q}}\right\rangle
>0
\end{equation}
holds in the thermodynamic limit
under the assumption~\eqref{nonpolarizedcondition}.
In other words, the many-body gap 
\begin{equation}
\Delta^{\ }_{\sigma,\bs{Q}}:=
\left\langle\Phi^{\ }_{\sigma,\bs{Q}}\left|
H
\right|\Phi^{\ }_{\sigma,\bs{Q}}\right\rangle
-
\left\langle\Psi^{\ }_{\sigma}\left|
H
\right|\Psi^{\ }_{\sigma}\right\rangle
>0
\end{equation}
is non-vanishing under the assumption~\eqref{nonpolarizedcondition}.

\subsection{
Proof of Eq.~(\ref{eq: lower bound spread})
           }

We are now going to work exclusively with the 
single-particle eigenstates
of the band Hamiltonian%
~(\ref{eq: def kinetic energy d}).

We start with the completeness relation
\begin{subequations}
\begin{equation}
\sum_{\bs{r}\in A}
\sum_{\alpha=A,B}\,
|\,\bs{r},\alpha \rangle\,\langle \alpha,\bs{r}\,|= 
\openone
\end{equation}
on the bipartite lattice $\Lambda=A\cup B$.
The normalized Bloch states $|\varphi^{\ }_{\bs{k},\sigma,\lambda}\rangle$ 
are defined in terms of the
single-particle eigenstates~$\chi^{\ }_{\bs{k},\sigma,\lambda,\alpha}$
by their overlaps
\begin{equation}
\langle \alpha,\bs{r}\,|\varphi^{\ }_{\bs{k},\sigma,\lambda}\rangle:=
\frac{
e^{+{i}\,\bs{k}\cdot\bs{r}}
     }
     {
\sqrt{N}
     }\,
\chi^{\ }_{\bs{k},\sigma,\lambda,\alpha}
\label{eq: overlap Bloch}
\end{equation}
for any given (Ising) spin $\sigma=\uparrow,\downarrow$ and band $\lambda=\pm$.
The normalization and phase factors are chosen so that
the orthonormality condition
\begin{equation}
\langle
\varphi^{\ }_{\bs{k},\sigma,\lambda}|
\varphi^{\ }_{\bs{k},\sigma,\lambda'}
\rangle=
\delta^{\ }_{\lambda,\lambda'}
\label{eq: orthonormality Bloch}
\end{equation}
with $\lambda,\lambda'=\pm$
holds for any given wavenumber $\bs{k}\in\mathrm{BZ}$ and any given
spin $\sigma=\uparrow,\downarrow$. The overlap%
~(\ref{eq: overlap Bloch})
is invariant under the translation
$\bs{k}\to\bs{k}+\bs{Q}$ 
for any $\bs{Q}$ that belongs to the reciprocal lattice of sublattice $A$
owing to the periodicity
\begin{equation}
\chi^{\ }_{\bs{k},\sigma,\lambda,\alpha}=
\chi^{\ }_{\bs{k}+\bs{Q},\sigma,\lambda,\alpha}.
\label{eq: periodicity chi}
\end{equation}
\end{subequations}

Wannier states are defined in terms of the Bloch states by
the unitary transformation
\begin{subequations}
\label{eq: def Wannier}
\begin{equation}
|\,\psi^{\ }_{\bs{z},\sigma,\lambda}\rangle:=
\frac{1}{\sqrt{N}}\,\sum_{\bs{k} \in \rm{BZ}}\,
e^{-{i}\,\bs{k}\cdot\bs{z}}\,
|\varphi^{\ }_{\bs{k},\sigma,\lambda}\rangle
\label{eq: def Wannier a}
\end{equation}
for any given unit cell of sublattice $A$
labeled by $\bs{z}$ and any given 
spin $\sigma=\uparrow,\downarrow$.
The orthonormality~(\ref{eq: orthonormality Bloch}) 
of the Bloch states thus carries over to the orthonormality
\begin{equation}
\langle\,\psi^{\ }_{\bs{z},\sigma,\lambda}
|\,\psi^{\ }_{\bs{z},\sigma,\lambda'}\rangle=
\delta^{\ }_{\lambda,\lambda'}
\label{eq: def Wannier b}
\end{equation}
of the Wannier states for any given unit cell labeled by $\bs{z}$ 
of sublattice $A$ and any given spin $\sigma=\uparrow,\downarrow$. 
For any $\lambda$ and $\alpha$,
the representation of the Wannier state on the  bipartite lattice
$\Lambda=A\cup B$ is the overlap
\begin{eqnarray}
\psi^{\ }_{\bs{r},\bs{z},\sigma,\lambda,\alpha}&:=&
\langle \bs{r},\alpha |\,\psi^{\ }_{\bs{z},\sigma,\lambda}\rangle
\nonumber\\
&=&
\frac{1}{\sqrt{N}}\,\sum_{\bs{k} \in \rm{BZ}}\,
e^{-{i}\,\bs{k}\cdot\bs{z}}\,
\langle \bs{r},\alpha |\varphi^{\ }_{\bs{k},\sigma,\lambda}\rangle
\nonumber\\
&=&
\frac{1}{N}\,\sum_{\bs{k} \in \rm{BZ}}\,
e^{+{i}\,\bs{k}\cdot(\bs{r}-\bs{z})}\,
\chi^{\ }_{\bs{k},\sigma,\lambda,\alpha}
\label{eq: def Wannier c}
\end{eqnarray}
for any given unit cell of sublattice $A$ labeled by $\bs{z}$ 
and any given spin $\sigma=\uparrow,\downarrow$.
The set of Wannier spinors
\begin{equation}
\psi^{\ }_{\bs{r},\bs{z},\sigma,\lambda}=
\left(\psi^{\ }_{\bs{r},\bs{z},\sigma,\lambda,\alpha}\right)
\label{eq: def Wannier d}
\end{equation}
resolves the identity since
\begin{equation}
\sum_{\bs{r}\in A}
\psi^{\dag}_{\bs{r},\bs{z} ,\sigma,\lambda }
\psi^{\   }_{\bs{r},\bs{z}',\sigma,\lambda'}
=
\delta^{\ }_{\bs{z},\bs{z}^{\prime}}\,
\delta^{\ }_{\lambda,\lambda^{\prime}}
\label{eq: def Wannier e}
\end{equation}
\end{subequations}
for any given spin $\sigma=\uparrow,\downarrow$.
We want to estimate the profile in space of the Wannier
states~(\ref{eq: def Wannier a}).

For that purpose, we consider the spread functional 
\begin{equation}
\label{eq: spread def}
\begin{split}
R^{(2)}_{\sigma,\lambda}:=&\,
\langle\,\psi^{\ }_{\bs{z}=0,\sigma,\lambda}\,|\,
\bs{r}^{2}
|\psi^{\ }_{\bs{z}=0,\sigma,\lambda}\,\rangle
\\
&-
|\,\langle\,\psi^{\ }_{\bs{z}=0,\sigma,\lambda}\,|\,
\bs{r}
|\psi^{\ }_{\bs{z}=0,\sigma,\lambda}\,\rangle\,|^{2}.
\end{split}
\end{equation}
Here, we are assuming, for simplicity, 
that there is only one band $\lambda$ that is occupied. 
If more bands are occupied,
we have to carry out a summation over all the occupied bands. 
Observe that by translational invariance $R^{(2)}_{\sigma,\lambda}$ 
is left unchanged under the global translation
$\bs{r},\bs{z}\to\bs{r}+\bs{R},\bs{z}+\bs{R}$ 
for any lattice vector $\bs{R}$.
Hence, the choice $\bs{z}=0$ in Eq.~\eqref{eq: spread def} can be
done without loss of generality.

We rewrite Eq.~\eqref{eq: spread def}, 
following Ref.~[\onlinecite{Thonhauser06}], as
\begin{subequations}
\begin{eqnarray}
R^{(2)}_{\sigma,\lambda}&:=&
R^{(2)}_{\rm{I}|\sigma,\lambda} 
+ 
\tilde{R}^{(2)}_{\sigma,\lambda},
\\
R^{(2)}_{\rm{I}|\sigma,\lambda}&:=&
\langle\,\psi^{\ }_{\bs{z}=0,\sigma,\lambda}\,|
\,\bs{r}^{2}
|\psi^{\ }_{\bs{z}=0,\sigma,\lambda}\,\rangle
\nonumber \\
&&-
\sum_{\bs{z}'}\,|\,
\langle\,\psi^{\ }_{\bs{z}=0,\sigma,\lambda}\,|
\,\bs{r}|\psi^{\ }_{\bs{z}',\sigma,\lambda}\,\rangle\,|^{2},
\\
\tilde{R}^{(2)}_{\sigma,\lambda}&:=&
\sum_{\bs{z}' \neq\bs{0}}\,
|\,\langle\,\psi^{\ }_{\bs{z}=0,\sigma,\lambda}\,|
\,\bs{r}
|\psi^{\ }_{\bs{z}',\sigma,\lambda}\,\rangle\,|^{2}.
\end{eqnarray}
\end{subequations}
Both $R^{(2)}_{\rm{I}|\sigma,\lambda}$ and $\tilde{R}^{(2)}_{\sigma,\lambda}$ 
are non-negative quantities 
that can be expressed in terms of $\bs{k}$-space summations as follows. 
First,
\begin{subequations}
\begin{widetext}
\begin{equation}
\begin{split}
R^{(2)}_{\rm{I}|\sigma,\lambda}&=
\sum_{j=x,y}\,
\frac{1}{N}
\sum_{\bs{k}\in\mathrm{BZ}}
\left[
\left(
\partial^{\ }_{k^{\ }_{j}}
\chi^{\dagger}_{\bs{k},\sigma,\lambda}
\right)
\left(
\partial^{\ }_{k^{\ }_{j}}
\chi^{\ }_{\bs{k},\sigma,\lambda}
\right)
-
\left(
\partial^{\ }_{k^{\ }_{j}}
\chi^{\dagger}_{\bs{k},\sigma,\lambda}\,\,
\chi^{\ }_{\bs{k},\sigma,\lambda}
\right) 
\left(
\chi^{\dagger }_{\bs{k},\sigma,\lambda}\,\,
\partial^{\ }_{k^{\ }_{j}}
\chi^{\ }_{\bs{k},\sigma,\lambda}
\right)
\right]
\\
&=
\sum_{j=x,y}\,
\frac{1}{N}
\sum_{\bs{k}\in\mathrm{BZ}}
\left[
\langle\,
\partial^{\ }_{k^{\ }_{j}}
\chi^{\ }_{\bs{k},\sigma,\lambda}\,|\,
\partial^{\ }_{k^{\ }_{j}}
\chi^{\ }_{\bs{k},\sigma,\lambda}
\rangle
-
\langle\,
\partial^{\ }_{k^{\ }_{j}}
\chi^{\ }_{\bs{k},\sigma,\lambda}\,|\,
\chi^{\ }_{\bs{k},\sigma,\lambda}
\rangle
\langle\,
\chi^{\ }_{\bs{k},\sigma,\lambda}\,|\,
\partial^{\ }_{k^{\ }_{j}}
\chi^{\ }_{\bs{k},\sigma,\lambda}\,
\rangle
\right]
\\
&=
\sum_{j=x,y}\,
\frac{1}{N}
\sum_{\bs{k}\in\mathrm{BZ}}
\langle\,
\partial^{\ }_{k^{\ }_{j}}
\chi^{\ }_{\bs{k},\sigma,\lambda}\,|\,
\left(
\openone 
- 
|\,\chi^{\ }_{\bs{k},\sigma,\lambda}\,\rangle\,
\langle\,\chi^{\ }_{\bs{k},\sigma,\lambda}|
\right)
|\,\partial^{\ }_{k^{\ }_{j}}
\chi^{\ }_{\bs{k},\sigma,\lambda}\,\rangle
\\
&\equiv
\sum_{j=x,y}\,
\frac{1}{N}
\sum_{\bs{k}\in\mathrm{BZ}}
\langle\,
\partial^{\ }_{k^{\ }_{j}}
\chi^{\ }_{\bs{k},\sigma,\lambda}\,|\,
Q_{\bs{k},\sigma,\lambda}
|\,\partial^{\ }_{k^{\ }_{j}}
\chi^{\ }_{\bs{k},\sigma,\lambda}\,\rangle
\\
&=
\frac{1}{N}
\sum_{\bs{k}\in\mathrm{BZ}}\, {\rm{tr}}
\left[ 
g^{\ }_{\bs{k},\sigma,\lambda}
\right],
\end{split}
\end{equation}
\end{widetext}
where
\begin{equation}
Q^{\ }_{\bs{k},\sigma,\lambda}:=
\openone 
- 
|\,\chi^{\ }_{\bs{k},\sigma,\lambda}\,\rangle\,
\langle\,\chi^{\ }_{\bs{k},\sigma,\lambda}|,
\end{equation}
is the single-particle projector operator on all the
Bloch states orthogonal to 
$|\,\chi^{\ }_{\bs{k},\sigma,\lambda}\,\rangle$,
the $2\times2$ matrix 
$
[
g^{\ }_{\bs{k},\sigma,\lambda}
]
$ 
has the components
\begin{equation}
g^{\ }_{\bs{k},\sigma,\lambda|i,j}:=
\mathrm{Re}
\left[\,
\langle\,
\partial^{\ }_{k^{\ }_{i}}
\chi^{\ }_{\bs{k},\sigma,\lambda}\,|\,
Q^{\ }_{\bs{k},\sigma,\lambda}
|\,\partial^{\ }_{k^{\ }_{j}}
\chi^{\ }_{\bs{k},\sigma,\lambda}\,\rangle
\right],
\end{equation}
\end{subequations}
labeled by the Euclidean indices $i,j\in\{x,y\}$ of two-dimensional
space, and tr denotes the trace over $i,j\in\{x,y\}$.
Second,
\begin{subequations}
\begin{equation}
\tilde{R}^{(2)}_{\sigma,\lambda}=
\sum_{j=x,y}\,
\frac{1}{N}\sum_{\bs{k}}\,
\left(
A^{\ }_{\bs{k},\sigma,\lambda|j}
-
\bar{A}^{\ }_{\sigma,\lambda|j}
\right)^2,
\end{equation}
where 
\begin{equation}
A^{\ }_{\bs{k},\sigma,\lambda|j}:= 
{i}\,
\langle\,\chi^{\ }_{\bs{k},\sigma,\lambda}|\,
\partial^{\ }_{k^{\ }_{j}}\chi^{\ }_{\bs{k},\sigma,\lambda}\,\rangle,
\end{equation}
and 
\begin{equation}
\bar{A}^{\ }_{\sigma,\lambda|j}:= 
\frac{1}{N}\,\sum_{\bs{k}}\,A^{\ }_{\bs{k},\sigma,\lambda|j}
\end{equation}
\end{subequations} 
denote the Berry connection and
the average of the Berry connection, respectively.
Under the gauge transformation 
\begin{equation}
|\chi^{\ }_{\bs{k},\sigma,\lambda}\,\rangle 
\rightarrow 
e^{{i}\varphi^{\ }_{\bs{k}}}\,|\chi^{\ }_{\bs{k},\sigma,\lambda}\,\rangle,
\end{equation}
it can be shown that
$R^{(2)}_{\rm{I}|\sigma,\lambda}$ remains invariant 
while $\tilde{R}^{(2)}_{\sigma,\lambda}$ does not. 

We will now establish a lower bound on the gauge invariant
quantity $R^{(2)}_{\rm{I}|\sigma,\lambda}$. 
To this end, we notice that, since $Q^{\ }_{\bs{k},\sigma,\lambda}$ 
is a projection operator with eigenvalues $0,1$, 
then, for any single-particle state $|\Psi^{\ }_{\bs{k},\sigma,\lambda}\rangle$, 
it follows that
\begin{equation}
\label{eq: inequality Psi}
0\leq 
\langle\,\Psi^{\ }_{\bs{k},\sigma,\lambda}|
Q^{\ }_{\bs{k},\sigma,\lambda}
|\Psi^{\ }_{\bs{k},\sigma,\lambda}\,\rangle.
\end{equation}
In particular, if we choose 
\begin{equation}
|\Psi^{\ }_{\pm,\bs{k},\sigma,\lambda}\,\rangle := 
|\partial^{\ }_{k_{x}}\chi^{\ }_{\bs{k},\sigma,\lambda}\,\rangle 
\pm
{i}
| \partial^{\ }_{k_{y}}\chi^{\ }_{\bs{k},\sigma,\lambda}\,\rangle,
\end{equation} 
the inequality~(\ref{eq: inequality Psi}) delivers
\begin{widetext}
\begin{equation}
\label{eq: inequal 2}
\mathrm{tr}\left[ g^{\ }_{\bs{k},\sigma,\lambda} \right]
\geq
\mp
{i}\,
\left(
\langle\,\partial^{\ }_{k^{\ }_{x}}\chi^{\ }_{\bs{k},\sigma,\lambda}\,
|\,\partial^{\ }_{k^{\ }_{y}}
\chi^{\ }_{\bs{k},\sigma,\lambda}\,\rangle
-
\langle\,\partial^{\ }_{k^{\ }_{y}}\chi^{\ }_{\bs{k},\sigma,\lambda}\,
|\,\partial^{\ }_{k^{\ }_{x}}
\chi^{\ }_{\bs{k},\sigma,\lambda}\,\rangle
\right).
\end{equation}
Summing~(\ref{eq: inequal 2}) over $\bs{k}$ and taking the thermodynamic limit
$N \rightarrow \infty$, delivers
\begin{equation}
\begin{split}
R^{(2)}_{\rm{I}|\sigma,\lambda}
=&\,
\frac{\mathcal{A}^{\ }_{c}}{(2\pi)^{2}}\int_{\rm{BZ}}\,d^{2}\bs{k}\,
\mathrm{tr}\left[ g^{\ }_{\bs{k},\sigma,\lambda} \right]
\\
\geq&\,
\pm \frac{\mathcal{A}^{\ }_{c}}{2\pi}\int_{\rm{BZ}}\,
\frac{d^{2}\bs{k}}{2\pi {i}}\,
\Big[
\langle\,\partial^{\ }_{k^{\ }_{x}}\chi^{\ }_{\bs{k},\sigma,\lambda}\,
|\,\partial^{\ }_{k^{\ }_{y}}
\chi^{\ }_{\bs{k},\sigma,\lambda}\,\rangle
-\langle\,\partial^{\ }_{k^{\ }_{y}}\chi^{\ }_{\bs{k},\sigma,\lambda}\,
|\,\partial^{\ }_{k^{\ }_{x}}
\chi^{\ }_{\bs{k},\sigma,\lambda}\,\rangle
\Big]
\\
=&\,
\pm
\frac{\mathcal{A}^{\ }_{c}}{2\pi}\,C^{\ }_{\sigma,\lambda},
\end{split}
\end{equation}
\end{widetext}
where $\mathcal{A}^{\ }_{c}$ is the area of the unit
cell of sublattice 
$A$ and  $C^{\ }_{\sigma,\lambda}$ are the
spin- and band-resolved Chern numbers.
Since $R^{(2)}_{\rm{I}|\sigma,\lambda} \geq 0$, it then follows
\begin{equation}
R^{(2)}_{\rm{I}|\sigma,\lambda}
\geq
\frac{\mathcal{A}^{\ }_{c}}{2\pi}\,|C^{\ }_{\sigma,\lambda}|,
\label{eq: lower bound proved}
\end{equation}
as a lower bound, proportional to the band Chern number, 
on the gauge invariant part of spread of the Wannier states%
~(\ref{eq: def Wannier a}).
It was shown in Ref.[~\onlinecite{Thonhauser06}] that, 
in the topological phase of the Haldane model~\cite{Haldane88},
$R^{(2)}_{\rm{I}|\sigma,\lambda}$ 
is finite while $\tilde{R}^{(2)}_{\sigma,\lambda}$ displays a logarithmic
divergence which is related to the non-zero Chern number of the bands. 
\section{
Competing Phases in the 
Hubbard model
        }

\begin{figure}
\includegraphics[width=0.48\textwidth]{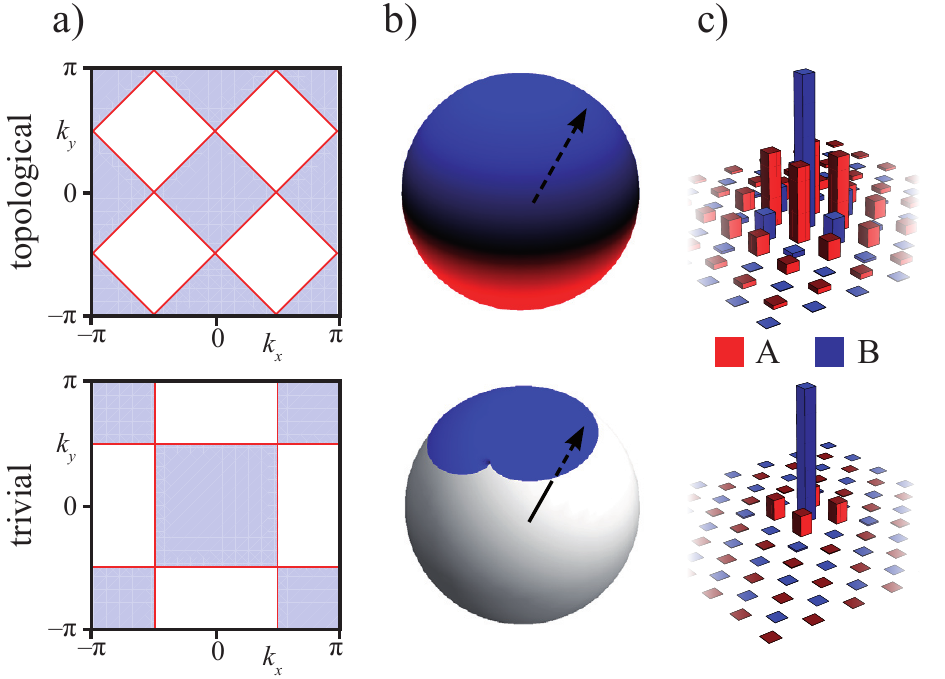}
\caption{ (Color online)
Comparison of the topological 
(top; $t^{\ }_{2}=1/\sqrt{2}$, $\mu^{\ }_\mathrm{s}=0$) 
and non-topological 
(bottom; $t^{\ }_{2}=0$, $\mu^{\ }_\mathrm{s}=1/\sqrt{2}$) 
single-particle model.
(a) The shaded area represents the Fermi see of the lower band 
at the commensurate filling fraction $N^{\ }_{\mathrm{e}}=N$
when $\kappa <1$.
(b) The eigenspinor $\chi^{\ }_{\bs{k},\sigma}$, 
when interpreted as a point on the surface of the unit sphere, 
swipes out the full surface of this sphere 
(a small portion of this sphere near one pole) 
as $\bs{k}$ takes values everywhere in the BZ for the topological 
(non-topological) band structure. 
(c) The spread of the Wannier states~\eqref{eq: def Wannier c} 
in real space indicates their
delocalized (localized) character for the topological (non-topological) 
band structure.
\label{fig:0} 
        }
\end{figure}

Any translationally invariant electronic single-particle Hamiltonian 
is described in momentum space by the dispersions and 
Bloch eigenfunctions of its bands.
Given the hopping amplitudes of a tight-binding model in 
position space, the dispersions and Bloch eigenfunctions
are uniquely determined. 
Once interactions are included, the many-body ground state of the model 
depends crucially on the properties of the 
dispersions (bandwidths; shapes of Fermi surface, e.g., nesting; ...)
and the properties of the Bloch states or Wannier states
(Chern numbers, localization in position space, ...).
The power of our approach, 
the flattening of the band with the factor $w^{\ }_{\bs{k}}$
from Eq.~(\ref{eq: def kinetic energy}),
is to disentangle the properties of the dispersions 
from those of the spinors to some extend
and to study their effects individually.

\subsection{Possible phases}

Let us list the anticipated ground states
for some limiting properties of the spinors and the dispersions
for $N^{\ }_e=N$ electrons when the two pairs of bands are separated
by a gap in the absence of interactions.

\begin{figure}[t]
\includegraphics[width=0.48\textwidth]{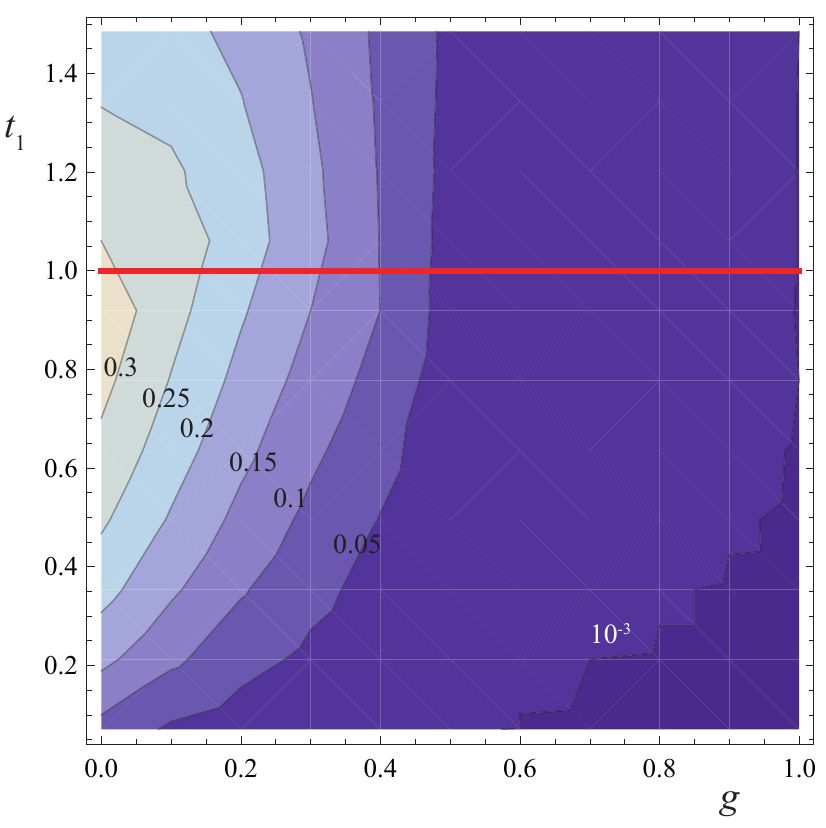}
\caption{ (Color online)
Numerical exact diagonalization results for flat bands $\kappa=1$ 
for the system size $L^{\ }_{x}=L^{\ }_{y}=25$
at the commensurate filling fraction $N^{\ }_{e}=N$.
Plotted is the value of 
the many-body excitation gap $\Delta/U$
between the energy of the fully spin polarized state $S=N$
and the lowest energy state in the sector with one spin flipped $S=N-2$
as a function of 
$g:=(2/\pi)\mathrm{arctan}|\mu^{\ }_{\mathrm{s}}/t^{\ }_{2}|$ 
and the nearest-neighbor hopping parameter $t^{\ }_1$.
The thick red line corresponds to the parameter choice made for  the 
plot in Fig.~1 of the Letter.
As discussed in the text, the spinors of the Bloch band become
fully sublattice polarized in the limit $t^{\ }_1\to0$
and the ferromagnetic ground state becomes degenerate in energy
with states from other spin sectors ($\Delta\to 0$).
\label{fig:1app} 
        }
\end{figure}

\begin{enumerate}
\item
\emph{(Almost) flat band and sublattice-polarized spinors.}
If the band is completely flat ($\kappa=1$), 
the ground state is macroscopically 
degenerate in the absence of interactions.
Moreover, when the Bloch states are fully sublattice polarized, 
i.e., when $\chi^{\dagger}_{\bs{k}}=(1,0)$ say, then
the kinetic energy becomes diagonal in position space 
[$\psi^{\dag}_{\bs{r},\bs{z},\sigma,-}=\delta^{\ }_{\bs{r},\bs{z}}\,(1,0)$].
Hence, any state with no less and no more 
than one electron per site on sublattice $A$ 
will be a ground state of the Hamiltonian, 
regardless of the spin-configuration of the electrons. 
Therefore, the ground state remains macroscopically degenerate, 
in spite of the presence of the Hubbard interaction.
This degeneracy can be lifted by introducing a small but finite bandwidth,
while keeping the full sublattice polarization of the spinors intact. 
Then, the low-energy degrees of freedom of the
model map on the conventional one-band Hubbard model at half filling with 
$t\ll U$, i.e., on the $t$-$J$ model.
An antiferromagnetic ground state will be selected.
An alternative way to lift the macroscopic degeneracy is 
to add longer range interactions, in which case other 
ground states may be stabilized~\cite{Neupert11b}.
Note that the fully sublattice polarized spinors discussed 
here imply that the  non-interacting model is topologically trivial.
\label{case 1}
\item
\emph{Flat band and not sublattice polarized spinors.}
As was shown in the Letter, 
if the spinors are sufficiently spread between the two sublattices,
i.e., if condition~\eqref{nonpolarizedcondition} is met, 
the ground state in the limit of flat bands 
is an Ising ferromagnet with full spin polarization.
\label{case 2}
\item
\emph{Strongly dispersing unnested band.}
For a strongly dispersing band without a nested Fermi surface, 
the Fermi liquid ground state is expected to be stable against 
sufficiently small repulsive Hubbard interactions. 
One generic instability at \emph{finite} interaction strength is 
Stoner ferromagnetism which, in contrast to the flat band ferromagnetism, 
does not automatically yield full spin polarization.
\label{case 3}
\item
\emph{Strongly dispersing nested band.}
If the Fermi surface is nested, arbitrary small Hubbard interactions
destabilize the Fermi liquid ground state toward an antiferromagnet.
\label{case 4}
\end{enumerate}

\begin{figure}[t]
\includegraphics[width=0.48\textwidth]{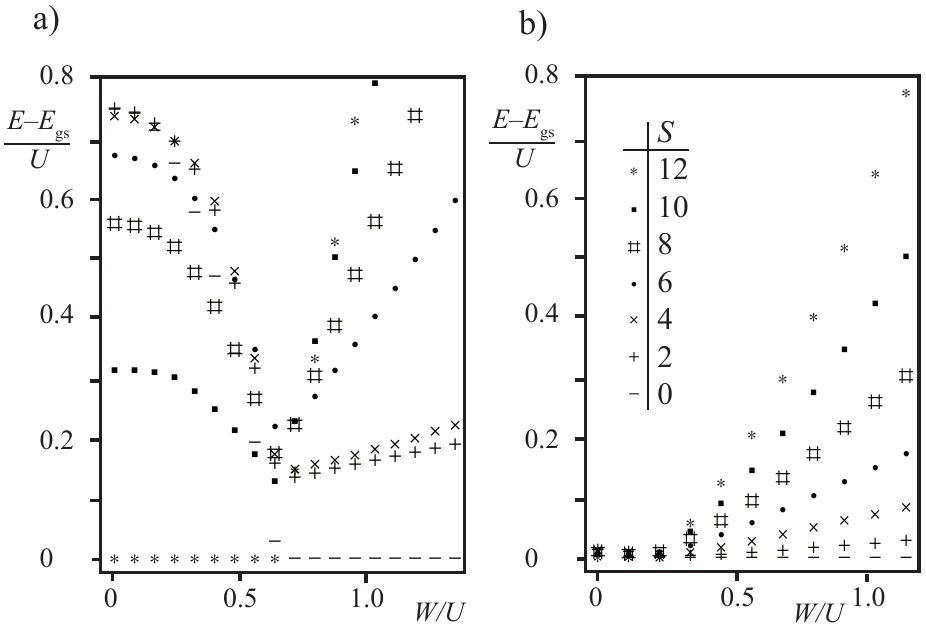}
\caption{ 
Numerical exact diagonalization results at 
the commensurate filling fraction $N^{\ }_{\mathrm{e}}=N$
as a function of the bandwidth $W$
for $L^{\ }_{x}=3,\ L^{\ }_{y}=4$. Plotted is the energy of the lowest state 
in different sectors of total spin $S$ (in units of $\hbar/2$) 
measured with respect to the ground state energy.
(a) Topological phase with 
$\mu^{\ }_{\mathrm{s}}=0$, $t^{\ }_{2}=1/\sqrt{2}$. 
The ground state is gaped and fully spin-polarized for $W/U<0.7$, 
while it is unpolarized for $W/U>0.7$.
(b) Topological trivial phase with 
$\mu^{\ }_{\mathrm{s}}=1/\sqrt{2}$, $t^{\ }_{2}=0$. 
The unpolarized ground state appears already for
very small values of $W/U$.
\label{fig:2app} 
        }
\end{figure}

\subsection{Realization in our model}

The shape of the Wannier states is controlled by the parameters 
$t^{\ }_2$ and $\mu^{\ }_{\mathrm{s}}$ in our model.
If they are topologically non-trivial ($|t^{\ }_2/\mu^{\ }_{\mathrm{s}}|>1$),
the Wannier states are relatively wide spread over several lattice sites
as illustrated in Fig.~\ref{fig:0} (c) and by Eq.%
~(\ref{eq: lower bound proved}).
In contrast, deep in the topological trivial regime 
$|t^{\ }_{2}/\mu^{\ }_{\mathrm{s}}|\ll1$, 
the Wannier states become more and more localized. 
Intuitively, their localization is more pronounced 
the smaller the area covered by the spinor $\chi^{\ }_{\bs{k},\sigma,\lambda}$
on the unit sphere as $\bs{k}$ takes values everywhere in the BZ 
[Fig.~\ref{fig:0} (b)].

In our model, the full sublattice polarization
is achieved by taking the limit
$t^{\ }_1\to 0$ while $|t^{\ }_2/\mu^{\ }_{\mathrm{s}}|<1$ 
(this limit is not shown in the Figures of the Letter). 
In the case of flat bands, this limit is the transition from 
phase~\ref{case 2} to phase~\ref{case 1} from the list above.
As shown in Fig.~\ref{fig:1app}, the many-body gap $\Delta$ of the 
flatband ferromagnetism collapses regardless of the value taken 
by the ratio 
$|t^{\ }_{2}/\mu^{\ }_{\mathrm{s}}|$ 
as $t^{\ }_{1}\to 0$.
The lowest energy states of each spin sector become degenerate
in this limit as the full $SU(2)$ spin-1/2 symmetry of the Hubbard
interaction is recovered in the limit $t^{\ }_{1}\to 0$.

We studied the effect of finite band width both in the 
topological trivial ($|t^{\ }_2/\mu^{\ }_{\mathrm{s}}|<1$) 
and non-trivial sector
($|t^{\ }_2/\mu^{\ }_{\mathrm{s}}|>1$) sector (Fig.~\ref{fig:2app}).
In both cases,
an interpolation between the phases~\ref{case 2} 
and~\ref{case 4} follows from changing the ratio $W/U$ 
[Fig.~\ref{fig:0}(a) illustrates the nesting of the Fermi surfaces for the
parameters chosen.].
The difference between topological non-trivial and trivial case 
is quantitative: While the many-body gap $\Delta$ protects the 
Ising ferromagnetic state against the effect of a sizable band width
in the former case, the transition to the antiferromagnetic state 
occurs already for very small $W/U$ in the latter.

\end{document}